\documentclass[pdflatex,sn-mathphys-num]{sn-jnl}

\usepackage[utf8]{inputenc}
\usepackage{mathrsfs}
\usepackage{graphicx}%
\usepackage{multirow}%
\usepackage{amsmath,amssymb,amsfonts}%
\usepackage{amsthm}%
\usepackage{mathrsfs}%
\usepackage[title]{appendix}%
\usepackage{textcomp}%
\usepackage{manyfoot}%
\usepackage{booktabs}%
\usepackage{algorithm}%
\usepackage{algorithmicx}%
\usepackage{algpseudocode}%
\usepackage{listings}%
\usepackage{longtable} 
\usepackage{booktabs} 
\usepackage{array}
\usepackage{tabularx}
\usepackage{float}
\theoremstyle{thmstyleone}%
\theoremstyle{thmstyletwo}%

\theoremstyle{thmstylethree}%

\begin{document}

\title[Article Title]{\bf \large Women in Theoretical Quantum Physics in Brazil: {\large \it demographics, career profiles, recognition, and leadership}}


\author*[1]{\fnm{Tatiana} \sur{Pauletti}}\email{tatiana.pauletti@unesp.br}

\author[2]{\fnm{Paula} \sur{Homem-de-Mello}}\email{paula.mello@ufabc.edu.br}

\author[3]{\fnm{Thereza } \sur{Paiva}}\email{tclp@if.ufrj.br}

\author*[1]{\fnm{Vivian V. } \sur{França}}\email{vivian.franca@unesp.br}

\affil*[1]{\orgdiv{Department of Physics and Mathematics}, \orgname{Institute of Chemistry, São Paulo State University (UNESP)}, \city{Araraquara}, \state{São Paulo}, \country{Brazil}}

\affil[2]{\orgdiv{Center of Natural Sciences and Humanities}, \orgname{Federal University of ABC},  \city{Santo André}, \state{São Paulo}, \country{Brazil}}

\affil[3]{\orgname{Instituto de Física, Universidade Federal do Rio de Janeiro (UFRJ)}, \city{Rio de Janeiro}, \state{Rio de Janeiro}, \country{Brazil}}


\abstract{Gender imbalance in Physics remains a persistent global challenge, and Brazil is no exception. While women account for only 24\%  of Physics faculty in the country, their representation in Quantum Physics is even smaller. In this work, we provide the first comprehensive overview of women working in Theoretical Quantum Physics in Brazil, here referred to as the SheQ (She + Quantum) community. Using data from the CNPq Lattes platform, we identify 93 researchers and analyze their geographic distribution, academic trajectories, scientific productivity, international experience, recognition through awards and fellowships, and engagement with initiatives promoting gender equity. Our results reveal both progress and persistent disparities: SheQ researchers have a strong scientific output, leadership roles, and international training; yet, their recognition through productivity fellowships remains modest, and their involvement in gender-related initiatives, although increasing among younger generations, remains limited. By combining quantitative indicators with institutional perspectives, we highlight structural barriers as well as opportunities for fostering a more inclusive environment in Quantum Physics. This study thus contributes to a broader reflection on how diversity not only promotes fairness but also strengthens creativity, innovation, and scientific progress.}


\maketitle

\section{Introduction}\label{sec1}


Gender balance in Physics has always been a challenging goal, not only in Brazil but worldwide \cite{MANIFEST,SheFigures2021,DPG2024,IOP2023,RSEF2024,Atominnen2024,JANU2024,Ipsos2024,DossieMulheresCiencia2025}. Studies consistently indicate that, from the early stages of schooling, girls tend to show less interest in mathematics, physics, chemistry, and engineering compared to boys \cite{Wang2016,oecd2015abc,DossieMulheresCiencia2025}. For a long time, this phenomenon was incorrectly attributed to supposed innate differences between male and female brains \cite{Meynell2013,Borthwick2021}. Modern neuroscience and education research have demonstrated that such claims are unfounded  \cite{Spelke2005,Nosek2009,Hyde2014,Lindgren2010,Chan2022,Gualtierotti2025,Rippon2023}: the disparities are not biological inevitabilities but rather the result of social and cultural conditioning. The way society, families, and schools encourage or discourage certain interests in children --- often unconsciously --- plays a decisive role in shaping career choices \cite{Lazzarini2018,Bello2022,Costa2024}. Girls are frequently steered toward fields perceived as “caring” or “human-centered,” while boys are encouraged to explore areas linked to technology, problem-solving, and abstraction \cite{Velho1998,Leta2003,Gomes2022}.

Over the past decades, numerous initiatives have sought to reverse this trend \cite{UcheNweje2025,IUPAP2022,MANIFEST,DossieMulheresCiencia2025}, aiming to attract more women into Physics, therefore creating an environment in which diversity is valued as a driver of creativity, innovation, and scientific progress. The participation of women brings not only different perspectives and approaches to problem-solving but also strengthens the collaborative and multidisciplinary nature of modern physics research. Despite these efforts, gender imbalances remain significant, particularly in highly specialized and competitive areas such as Quantum Physics \cite{MANIFEST}.

In this work, we focus on mapping the current landscape of women working in Theoretical Quantum Physics in Brazil. For clarity and identity, we refer to them as SheQ --- a term coined from “She” and “Quantum.” Our study seeks to provide a comprehensive overview of this group by examining several dimensions: {\it i)} the overall gender distribution within the field, {\it ii)} the geographic distribution of SheQ across Brazil’s regions, {\it iii)} the typical academic and professional profiles of SheQ, {\it iv)} the forms of recognition they receive within the Physics community, and {\it v)} engagement in equality initiatives and leadership roles influencing institutional policies. By combining statistical data with qualitative insights, we aim to shed light on both the progress achieved and the structural barriers that still need to be addressed, offering perspectives for fostering a more equitable and inclusive future in Quantum Physics in Brazil.

\section{Representation of Women in Brazilian Physics}

\begin{figure}[ht]
    \centering
    \includegraphics[width=0.9\linewidth]{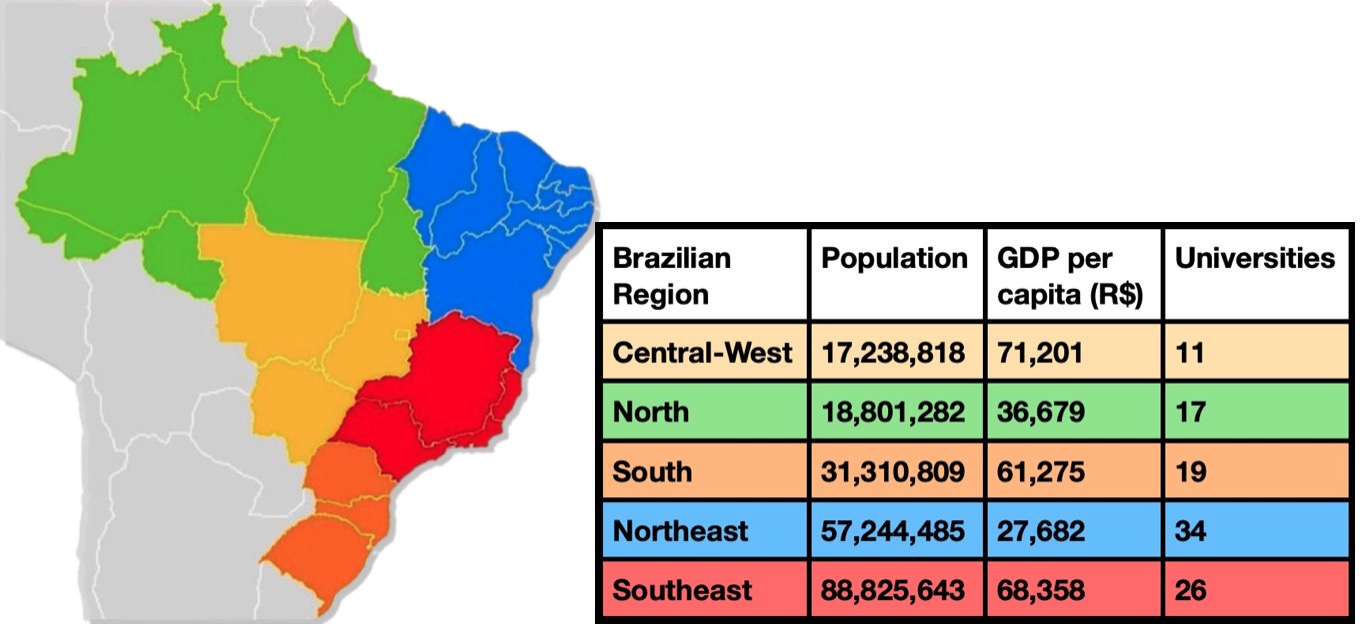}
    \caption{Brazilian regions \cite{mapa_brasil} and their population estimates \cite{ibge2025populacao}, Gross Domestic Product (GDP) per capita \cite{ibge2025pib} and total number of public (Federal and State) Universities \cite{andifes2025_69uf,projetosol_universidades_estaduais}.}
    \label{fig:brazilian_regions}
\end{figure}

Brazil’s five macro-regions --- North, Northeast, Southeast, South, and Central-West --- form a heterogeneous academic landscape, with substantial variation in population, economic indicators, and the number of public universities, as shown in Figure \ref{fig:brazilian_regions}. The national research ecosystem is sustained primarily by the National Council for Scientific and Technological Development (CNPq) and by the Brazilian Federal Agency for Support and Evaluation of Graduate Education (CAPES), which provide core funding for graduate training, research fellowships, and institutional support across public Universities. These efforts are reinforced by state research foundations (FAPs, such as FAPESP, FAPEMIG, FAPERJ, and others) which enhance local scientific capacity and help mitigate regional disparities.

\begin{figure}[ht]
\centering
\includegraphics[width=0.8\linewidth]{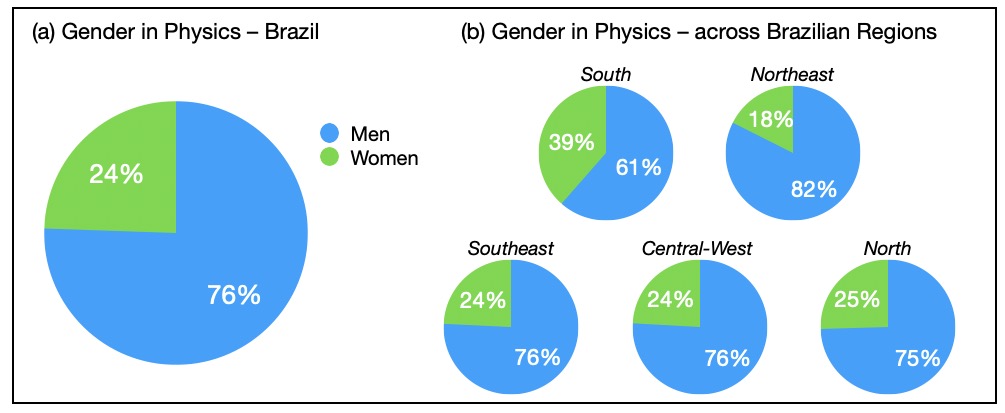}
\caption{\label{fig1}Gender imbalance in Physics: (a) throughout Brazil and (b) distribution by geographic region. The data were retrieved from the CNPq databases \cite{cnpq_lattes} on July 8, 2025, according to the filters listed in the Appendix.}\label{fig0}
\end{figure}

Recent data from CNPq \cite{cnpq_lattes,cnpq_lattes_plataforma,cnpq_chamada18_2024} provide a snapshot of the composition of Physics faculty in Brazil, encompassing 3,116 researchers working at federal or state universities and research institutes.  As illustrated in Figure \ref{fig1}a, women represent only 24\%  of this group.  Focusing on the geographic distribution, Fig.~\ref{fig1}b shows that the Southeast, Central-West, and North regions exhibit a gender imbalance similar to the national average, while the South stands out with the best gender balance, with women representing 39\% of faculty. In contrast, the Northeast shows the greatest imbalance in Physics, with only 18\% of women academics.

\begin{figure}[ht]
\centering
\includegraphics[width=0.98\linewidth]{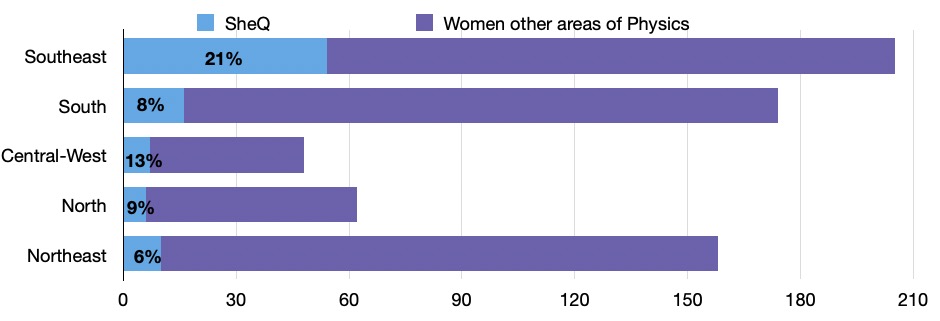}
\caption{\label{fig2} Women in Physics: Distribution of SheQ and other areas across Brazil’s geographic regions. The figure also shows the percentage of female researchers in each region who belong to the SheQ community. }
\end{figure}

Among the 762 women physicists in Brazil, 93 constitute the SheQ community (see Appendix) --- those working specifically in Theoretical Quantum Physics --- representing 12\% of all women in Physics, a considerable share within the field. Nevertheless, they account for only 4\% of all Physics faculty nationwide. Figure \ref{fig2} presents the total numbers across Brazil’s geographic regions, along with the proportion of women in each region active in Theoretical Quantum Physics.

\begin{figure}[ht]
\centering
\includegraphics[width=0.95\linewidth]{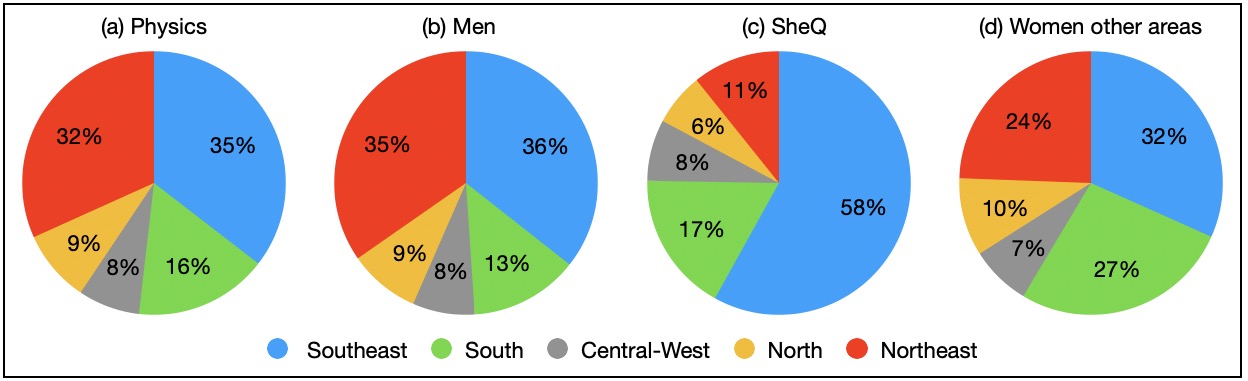}
\caption{\label{fig3}Geographic distribution: (a)  Physics community, (b) male researchers, (c) SheQ community, and (d) women in other areas of Physics. }
\end{figure}

Figures~\ref{fig3}a and \ref{fig3}b show that both men and the overall population of researchers are distributed almost equally between the South/Southeast and the Central-West/North/Northeast regions, with only small differences, as expected, since men comprise the majority of the total. In contrast, SheQ researchers display a pronounced regional concentration: 75\% are affiliated with institutions in the South and Southeast (Fig.~\ref{fig3}c), compared to just 59\% of women in other areas of Physics (Fig.~\ref{fig3}d). 

The concentration of SheQ researchers in Southeast and South may be associated with the greater availability of financial resources in these regions. A recent data report \cite{mcti_dispendios_estaduais} indicates that, on average between 2010 and 2023, state government investments in Science and Technology were distributed as follows: 68\% in the Southeast, 12\% in the South, 6\% in the Central-West, 4\% in the North, and 11\% in the Northeast. 

\section{SheQ Brazil --- profiles, recognition, and leadership}

We now focus exclusively on the average profile of SheQ researchers (Appendix), aiming to gain a deeper understanding of the community’s composition, characteristics, and patterns. 

\begin{figure}[ht]
\centering
\includegraphics[width=0.6\linewidth]{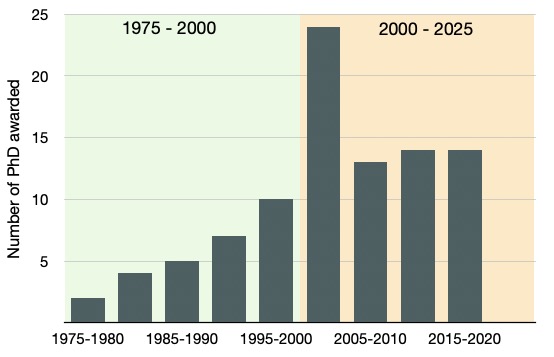}
\caption{ Periods in which SheQ researchers obtained their PhDs, grouped in five-year intervals. Data from CNPq \cite{cnpq_lattes_plataforma}.}
\label{fig:figure4}
\end{figure}

Figure \ref{fig:figure4} shows SheQ researchers according to the periods in which they completed their PhDs, grouped into five-year intervals, providing insight into the community’s generational trends. Between 1975 and 2000, 28 SheQ members earned their PhDs. In the next 25 years, this number more than doubled to 65, meaning that two out of every three SheQ researchers obtained their doctorates in the last quarter-century --- consistent with the recent growth of the quantum physics field and increased investment in it. Interestingly, none of the SheQ researchers have completed their PhDs in the past five years, likely reflecting the typical gap between finishing a doctorate and securing a permanent position, as well as possible disruptions caused by the COVID-19 pandemic. However, it is important to note that career advancement in science depends on judgments made by more senior colleagues --- judgments that often contain arbitrary and subjective elements disadvantaging women, as recently discussed \cite{nas2007}. Implicit and often unacknowledged gender bias has historically shaped evaluation processes, producing cumulative effects that restrict women’s access to prestige, promotions, and honors, and may partly explain delays or difficulties in their advancement to permanent and leadership positions.

\begin{table}[ht]
\caption{\label{tab1}Average profile of a SheQ researcher (Appendix).}
\begin{tabular}{r p{6cm}}
\hline
\toprule%
 44 & Scientific articles\\
13 & First author	\\
	8& Last author \\\hline
    \toprule%
	12 & Undergraduate students supervision \\
	8 & Graduate students supervision \\\hline
        \toprule%
	9 & Research projects coordination \\
40 & Participation in Scientific Conferences	\\
	7 & Organization of Scientific Conferences\\\hline
\toprule%
     \hspace{0.8cm}24 months & Months abroad as Postdoctoral fellow $^{(a)}$	\\
	1 & Registered patents$^{(b)}$ \\
5 & Awards and recognitions$^{(c)}$\\\hline
\toprule%
\end{tabular}
{\footnotesize ${(a)}$ average for the 48 SheQ with postdoctoral experience abroad;\\ ${(b)}$ average for the 8 SheQ with at least one registered patent;\\ ${(c)}$ average for the 54 SheQ who have received an award.}
\end{table}

\begin{figure}[b!]
\centering
\includegraphics[width=0.95\linewidth]{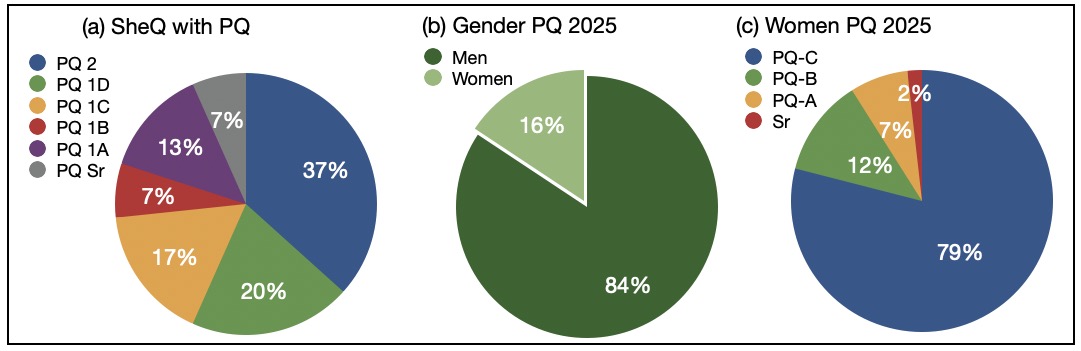}
\caption{\label{fig5} (a) SheQ recognition through CNPq Productivity Fellowships (PQ) across the old (prior to 2025) categories. (b) Gender balance within the PQ 2025. (c) Women's distribution within the new categories within the PQ 2025. Data from CNPq \cite{cnpq_lattes_plataforma,cnpq_bolsas_curso,cnpq_chamada18_2024}}
\end{figure}

Table \ref{tab1} summarizes the average profile of a typical SheQ researcher, including scientific productivity, student supervisions, project coordinations, participation in scientific conferences, international experience, and awards. On average, a SheQ researcher has authored 44 scientific papers, being the first author in $29\%$ and the last author in $18\%$ of them --- an indication of her established leadership within her research group. This trend becomes clearer when analyzing the temporal evolution over the past 15 years, in Figure \ref{fig5}. Although year-to-year fluctuations are observed (Fig. \ref{fig5}a), in particular for the number of students supervised, when data are grouped into a five-year window (Fig. \ref{fig5}b),  the share of papers as first author decreases over time, while the proportion as last author increases. 

In terms of academic leadership, on average, a SheQ researcher has coordinated 9 research projects and supervised 12 undergraduate and 8 graduate students. Furthermore, 53\% of the SheQ community spent a period abroad as postdoctoral fellows, visiting researchers, or some other kind of sabbatical leave, with an average stay of 24 months, reflecting their strong international experience. In addition, SheQ researchers have participated in about 40 scientific conferences and contributed to the organization of 7.

While only 8\% of the SheQ have registered patents, about 59\% of the community have received at least one award or distinction for their scientific contributions, with an average of five awards per person among those recognized. We also examined the recognition of SheQ productivity through the CNPq Productivity Fellowships across different categories. 

\begin{figure}[ht]
\centering
\includegraphics[width=0.9\linewidth]{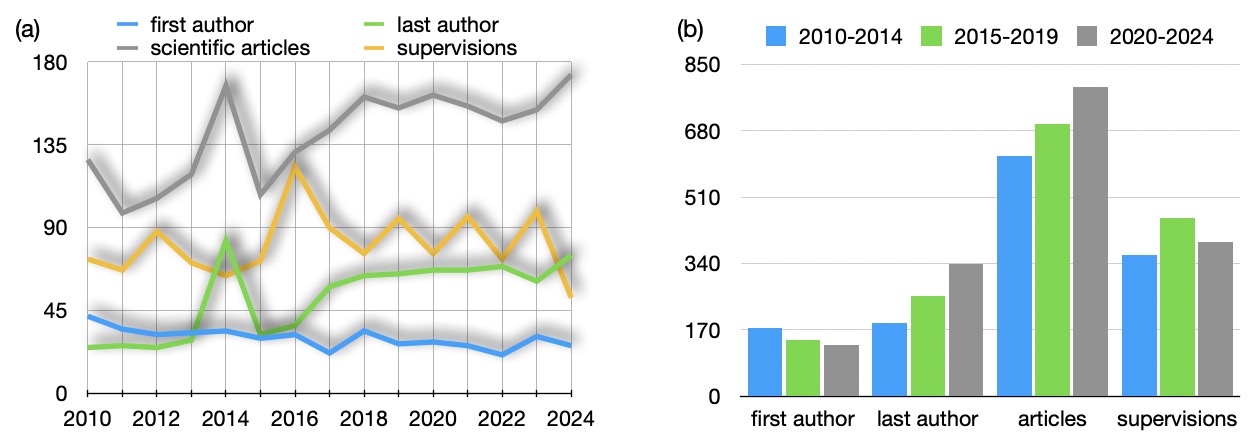}
\caption{\label{fig:temporal} Temporal evolution of SheQ researchers’ academic activities over the past 15 years: (a) year-to-year fluctuations and (b) five-year windows showing general trends. Data from CNPq \cite{cnpq_lattes_plataforma}.}
\end{figure}

Only 32\% of SheQ members hold a PQ fellowship, and more than half of them are concentrated in the two initial categories, as shown in Figure \ref{fig:temporal}a: 37\% in PQ 2 and 20\% in PQ 1D. Although the recognition via PQ within the SheQ community is still modest, it appears slightly better than that of women in the broader Physics community, where only 16\% of the fellowships were awarded to women, as shown in the most recent PQ results (Fig. \ref{fig:temporal}b).  Figure \ref{fig:temporal}c further indicates that the vast majority of these fellowships for women are concentrated in the two lowest categories, now denominated PQ-C and PQ-B, which together account for 91\%, with only 9\% distributed among the highest levels.

Beyond their scientific achievements, SheQ members also engage --- albeit to varying degrees --- in initiatives aimed at promoting gender equality in science. Overall, 53\% of them have participated in such activities, a proportion that may appear modest given that all have directly faced the consequences of gender imbalance throughout their careers.

A generational perspective, however, reveals subtle but meaningful shifts. Among those who obtained their PhDs in earlier years (1975--2000), 42\% have engaged in gender-related initiatives. In contrast, within the younger group (2001-2025), this share increases considerably, with 58\% getting involved. While the difference is not dramatic, it suggests a gradual increase in awareness and willingness to act among newer generations.

This evolving engagement opens a broader discussion on how women in physics perceive and respond to sexism in science. It also underscores the importance of institutional milestones, such as Prof. Thereza Paiva’s, leadership of the Advisory Committee of the Physics and Astronomy Area at CNPq since 2025, and Prof. Kaline Coutinho’s, leadership of the Physics and Astronomy Area at CAPES since 2024. Their voices are particularly meaningful here: Prof. Paiva offers her perspective on the significance and responsibilities of occupying this position at CNPq, while Prof. Coutinho highlights how her coordination at CAPES has enabled the incorporation of concrete measures to reduce gender disparities and to support early-career scholars:

\begin{quote}
      \begin{flushright}\qquad {\it ``Over the past years, it has been noticeable that women are actively seeking to occupy leadership roles in science, particularly in instances where science policy is discussed and implemented. The Advisory Committee of the Physics and Astronomy Area at CNPq oversees the distribution of research grants and research funds for Physics and Astronomy. Gender related issues, such as motherhood and its impact on academic productivity, must be taken into account. Having a woman as a coordinator not only reinforces the need to address these gender related issues, but serves as an example for younger scientists that these roles are also achievable by women."}\\
      \\
    {\bf Prof. Thereza Cristina de Lacerda Paiva, UFRJ}
    \end{flushright}
\end{quote}

\begin{quote}
  \begin{flushright}
    \qquad {\it ``I agree with Thereza's remarks. Furthermore, I would like to add that my role in the Coordination of Physics and Astronomy field at CAPES contributed to the inclusion of female faculty (with children under seven years old) and young faculty (with less than 10 years after their PhD degree and four years of being hired) in the graduate programs of the field. These faculty will be exempt from the requirement to be equally involved in all scientific activities and production. This initiative should not be understood as an incentive to underproductivity, but rather as a necessary measure to reduce pressure and allow these faculty to develop academically during these critical phases of their careers. Support for both categories of graduate faculty was introduced based on CAPES data, which reveals that only 16\% of the faculty in graduate programs in the field are women and that, in fact, there is a decline in the average scientific production of women who have children within a seven-year period. The same analysis showed that fatherhood does not affect the productivity of men. Another concerning statistic indicates that only 12\% of graduate faculty are under 40. Therefore, based on CAPES data, I sought to promote a management approach focused on improving the overall quality of the graduate education while also addressing gender inequality, among other pressing challenges."} 
    \\
   \\ {\bf Prof. Kaline Rabelo Coutinho, USP}
    \end{flushright}
\end{quote}

Female representation in decision-making roles not only provides visible role models for younger generations but also ensures that gender perspectives are incorporated into institutional policies and priorities \cite{Armenteras2025}. Such representation, crucial for shaping a more equitable and inclusive scientific environment, is exemplified in the testimonies of Professors Paiva and Coutinho, who highlight both the persistent challenges faced by women in academic physics and the transformative impact of having women in key leadership roles within Brazil’s main science funding and evaluation institutions. As an inspiring example, at present, there is also a perfect gender balance in the Coordination of the Astronomy and Physics Area at São Paulo Research Foundation
(FAPESP), where three of its six members are women: Prof. Angela Cristina Krabbe (IAG), Prof. Rosângela Itri (IFUSP), and Prof. Vivian Vanessa França (UNESP). The presence of women in leadership roles gives us hope that unconscious bias will play a lesser role, opening the door to a reasonable expectation of reduced gender inequality (at least concerning gender, but let us hope also in other societal issues), and pointing to a more open and inclusive scientific community for everyone.

\section{Final Remarks}

Diversity in science is not merely a matter of fairness or representation; it is a catalyst for progress. Just as friction enables movement, allowing us to walk, generate heat, or produce light, the presence of diverse perspectives generates the intellectual “friction” that stimulates innovation. In homogeneous groups, ideas tend to circulate within familiar boundaries, reinforcing the status quo. By contrast, diversity --- whether of gender, background, or worldview --- creates opportunities for challenging assumptions, refining arguments, and developing novel approaches. In Physics, and particularly in Quantum Physics, where complex problems often demand unconventional solutions, this creative friction can be decisive. It compels us to be more inventive, to push beyond comfort zones, and to engage in deeper critical thinking. In this sense, diversity is not only an ethical imperative but also a strategic advantage for advancing scientific discovery.

\section*{Declarations}

\begin{itemize}
\item Funding: {\bf T. Pauletti} is supported by the National Council
of Technological and Scientific Development CNPq (grand number 140854/2021-5). 
{\bf P. Homem-de-Mello} is thankful to the Brazilian agencies CNPq (grant number
305381/2022-9), Instituto Nacional de Ciência e Tecnologia (INCT, Materials Informatics, grant number 371610/2023-0), and Financiadora de Estudos e Projetos (FINEP, grant numbers 0038/21 and
0288/22).
{\bf T. Paiva} acknowledges financial support from 
Funda\c{c}\~ao Carlos Chagas Filho de Amparo \`a Pesquisa do Estado do Rio de Janeiro (grant numbers
E-26/200.959/2022 and E-26/210.100/2023), and from CNPq (grant numbers 308335/2019-8, 403130/2021-2,  and 442072/2023-6). 
{\bf V. V. França} is supported by São Paulo Research
Foundation FAPESP (grant numbers 2021/06744-8, 2025/02935-4, and 2024/00998-6) and CNPq (grant
numbers 403890/2021-7, 306301/2022-9).
\item Conflict of interest/Competing interests (check journal-specific guidelines for which heading to use):
{The authors declare no competing interests.}
\item Ethics approval and consent to participate: {Not applicable}
\item Consent for publication: {Not applicable}
\item Data availability: {All data used in this study are publicly accessible from the CNPq database \cite{cnpq_bolsas_curso} and the Lattes Platform \cite{cnpq_lattes, cnpq_lattes_plataforma}. The specific curricula analyzed are listed in the Appendix.}
\item Materials availability: {Not applicable}
\item Code availability: {Not applicable}
\item Author contribution: {T. Pauletti collected, filtered, and curated the data. V. V. França organized the dataset and performed the initial interpretations. All authors contributed to the final interpretations and writing of the manuscript.}
\end{itemize}

\noindent


\begin{appendices}

\section{}\label{secA1}

\begin{enumerate}
\item {\bf 93 SheQ members}
The 93-SheQ list was extracted from the CNPq database \cite{cnpq_lattes_plataforma} in July 2025 using keywords related to quantum science, precisely: 
\begin{center}
computação quântica\\
correlações quânticas\\
emaranhamento quântico\\
fenômenos quânticos coletivos\\
física quântica\\
gases quânticos\\
materiais quânticos\\
matéria condensada\\
mecânica quântica\\
óptica quântica\\
sistemas fortemente correlacionados\\
sistemas quânticos\\
teoria quântica\\
informação quântica\\
transições de fase quânticas. 
\end{center}

The search returned 12,547 curricula, of which 2,707 belonged to women. The final list of 93 researchers was selected by applying the following filters: 

\noindent{\it i)} women holding permanent positions in higher education or research institutions, \\
{\it ii)} currently working in theoretical physics, and \\
{\it iii)} excluding those primarily focused on education or interdisciplinary areas (e.g., biophysics, physical chemistry). \\

\item \href{https://www.iq.unesp.br/#!/mulheres-na-ciencia/sheq-members/}{\textbf{SheQ List online:}} 
This list may not be exhaustive and, therefore, we will keep it online for future updates \cite{unesp_sheq}.\\

\item {\bf Figures 1–6 and Table 1 ---} data were retrieved from the Lattes Platform (Data and Statistics), filtered by: region, academic rank (Adjunct, Associate, Assistant, and Full Professor), major field (Exact and Earth Sciences), specific area (Physics), and country (Brazil), reflecting the update of July 8, 2025 \cite{cnpq_lattes}.\\

\item \href{https://www.iq.unesp.br/#!/mulheres-na-ciencia/iniciativas-no-brasil/}{\textbf{List of Initiatives to Promote Women in Science} ---} 
we have compiled a set of initiatives in Brazil, with the intention of continuously updating and expanding this database \cite{unesp_iniciativas}.

\end{enumerate}

\vspace{-0.1cm}
\newpage
\scriptsize
\newcommand{\lattes}[2]{\href{#1}{#2}}

\begingroup
\setlength{\LTleft}{\fill}
\setlength{\LTright}{\fill}
\begin{longtable}{@{}l >{\raggedright\arraybackslash}p{8cm}@{}}
\toprule
\textbf{SheQ Member} & \textbf{Institution}  \\
\midrule
\endfirsthead
\toprule
\textbf{SheQ Member} & \textbf{Institution}  \\
\midrule
\endhead
\bottomrule
\endfoot

\lattes{http://lattes.cnpq.br/7866377101423210}{Adriana Pedrosa Biscaia Tufaile} & Universidade de São Paulo \\
\lattes{http://lattes.cnpq.br/9274147713584765}{Alessandra Ferreira Albernaz} & Universidade de Brasília \\
\lattes{http://lattes.cnpq.br/7611599930837576}{Alessandra Nascimento Braga} & Universidade Federal do Pará \\
\lattes{http://lattes.cnpq.br/3500088234156539}{Ana Claudia Monteiro Carvalho} & Universidade Federal de Juiz de Fora \\
\lattes{http://lattes.cnpq.br/5900065640583722}{Ana Cristina Sprotte Costa} & Universidade Federal do Paraná \\
\lattes{http://lattes.cnpq.br/2863779279262031}{Ana Júlia Silveira Mizher} & Instituto de Física Teórica – UNESP \\
\lattes{http://lattes.cnpq.br/9802635232762865}{Andrea Brito Latge} & Universidade Federal Fluminense \\
\lattes{http://lattes.cnpq.br/3395423149193260}{Andreia Mendonça Saguia} & Universidade Federal Fluminense \\
\lattes{http://lattes.cnpq.br/4323567432559309}{Angela Burlamaqui Klautau} & Universidade Federal do Pará \\
\lattes{http://lattes.cnpq.br/7012069649140656}{Angela Foerster} & Universidade Federal do Rio Grande do Sul \\
\lattes{http://lattes.cnpq.br/6396256338493731}{Angsula Ghosh} & Universidade Federal do Amazonas \\
\lattes{http://lattes.cnpq.br/0180245906512843}{Arlene Cristina Aguilar} & Universidade Estadual de Campinas \\
\lattes{http://lattes.cnpq.br/0644505365952110}{Barbara Lopes Amaral} & Universidade de São Paulo \\
\lattes{http://lattes.cnpq.br/7598434387418698}{Belita Koiller} & Universidade Federal do Rio de Janeiro \\
\lattes{http://lattes.cnpq.br/6417216091455652}{Bertha María Cuadros Melgar} & Universidade de São Paulo \\
\lattes{http://lattes.cnpq.br/0237981341645476}{Celia Maria Alves Dantas} & Universidade Federal de Goiás \\
\lattes{http://lattes.cnpq.br/7889980304354260}{Daniela Szilard Le Cocq D'Oliveira} & Universidade Federal do Rio de Janeiro \\
\lattes{http://lattes.cnpq.br/0520649409009819}{Danuce Marcele Dudek} & Universidade Federal da Fronteira Sul \\
\lattes{http://lattes.cnpq.br/4521038966994688}{Debora Peres Menezes} & Universidade Federal de Santa Catarina \\
\lattes{http://lattes.cnpq.br/8953964693512884}{Denise da Costa Assafrão de Lia} & Universidade Federal do Espírito Santo \\
\lattes{http://lattes.cnpq.br/5334131860888930}{Diana Esther Tuyarot de Barci} & Instituto Federal de Minas Gerais \\
\lattes{http://lattes.cnpq.br/4958262952068040}{Dyana Cristine Duarte} & Universidade Federal de Santa Maria \\
\lattes{http://lattes.cnpq.br/8003910745506954}{Elena Konstantinova} & Instituto Federal de Minas Gerais \\
\lattes{http://lattes.cnpq.br/5486014223324523}{Eliane Pereira} & Universidade Federal do Sul e Sudeste do Pará \\
\lattes{http://lattes.cnpq.br/7058872675355251}{Erica de Mello Silva} & Universidade Federal de Mato Grosso \\
\lattes{http://lattes.cnpq.br/7024420193968007}{Erika de Carvalho Bastone} & Universidade Federal de São João del-Rei \\
\lattes{http://lattes.cnpq.br/7285683672412029}{Érika Dias Cabral} & Universidade do Estado do Rio de Janeiro \\
\lattes{http://lattes.cnpq.br/6885659314359356}{Gabriela Barreto Lemos} & Universidade Federal do Rio de Janeiro \\
\lattes{http://lattes.cnpq.br/0473784991529055}{Gabrielle Weber Martins} & Universidade de São Paulo \\
\lattes{http://lattes.cnpq.br/2248844091470333}{Ginette Jalbert de Castro Faria} & Universidade Federal do Rio de Janeiro \\
\lattes{http://lattes.cnpq.br/0114229092431284}{Giovana Trevisan Nogueira} & Universidade Federal de Juiz de Fora \\
\lattes{http://lattes.cnpq.br/2763504129786617}{Halyne Silva Borges} & Instituto Federal do Triângulo Mineiro \\
\lattes{http://lattes.cnpq.br/1857810385456326}{Hatsumi Mukai} & Universidade Estadual de Maringá \\
\lattes{http://lattes.cnpq.br/9602029911336683}{Helena de Souza Bragança Rocha} & Universidade de Brasília \\
\lattes{http://lattes.cnpq.br/9116604783066746}{Helena Maria Petrilli} & Universidade de São Paulo \\
\lattes{http://lattes.cnpq.br/4017025843870236}{Hilma Helena Macedo de Vasconcelos} & Universidade Federal do Ceará \\
\lattes{http://lattes.cnpq.br/6389450935104315}{Hiromi Iwamoto} & Universidade Estadual de Londrina \\
\lattes{http://lattes.cnpq.br/7512625230047947}{Isabel Tamara Pedron} & Universidade Estadual do Oeste do Paraná \\
\lattes{http://lattes.cnpq.br/3026771014535770}{Isabela Porto Cavalcante} & Universidade Federal de Mato Grosso do Sul \\
\lattes{http://lattes.cnpq.br/6549861827328519}{Jemima Pereira Guedes} & Universidade Federal do Recôncavo da Bahia \\
\lattes{http://lattes.cnpq.br/8599236000105857}{Jessica Edith Quispe Bautista} & Universidade Federal do Rio Grande do Norte \\
\lattes{http://lattes.cnpq.br/9205662588542783}{Kaline Rabelo Coutinho} & Universidade de São Paulo \\
\lattes{http://lattes.cnpq.br/6438122475771528}{Krissia de Zawadzki} & Universidade de São Paulo \\
\lattes{http://lattes.cnpq.br/7502758036614082}{Lara Kühl Teles} & Instituto Tecnológico de Aeronáutica \\
\lattes{http://lattes.cnpq.br/0481046611049749}{Letícia Faria Domingues Palhares} & Universidade do Estado do Rio de Janeiro \\
\lattes{http://lattes.cnpq.br/0781577612002688}{Letície Mendonça Ferreira} & Universidade Federal do ABC \\
\lattes{http://lattes.cnpq.br/3187273888989883}{Liliana Sanz de la Torre} & Universidade Federal de Uberlândia \\
\lattes{http://lattes.cnpq.br/9147986360105551}{Lídia Carvalho Gomes} & Universidade Federal de Pernambuco \\
\lattes{http://lattes.cnpq.br/3945403573894154}{Luana Sucupira Pedroza} & Universidade de São Paulo \\
\lattes{http://lattes.cnpq.br/8089333823109715}{Lucy Vitória Credidio Assali} & Universidade de São Paulo \\
\lattes{ http://lattes.cnpq.br/1574084702480992}{Malena Osorio Hor-Meyll} & Universidade Federal do Rio de Janeiro \\
\lattes{ http://lattes.cnpq.br/7018014202492783}{Margarida Maria Rodrigues Negrão} & Universidade Federal do Pampa \\
\lattes{ http://lattes.cnpq.br/1633363001210472}{Marcia Moutinho} & Universidade Estadual de Mato Grosso do Sul \\
\lattes{http://lattes.cnpq.br/3251079293038925}{Maria Beatriz de Leone Gay Ducati} & Universidade Federal do Rio Grande do Sul \\
\lattes{ http://lattes.cnpq.br/2376256043612364}{Maria Caballero Tijero} & Pontifícia Universidade Católica de São Paulo \\
\lattes{ http://lattes.cnpq.br/3936407003118409}{Maria Carolina de Oliveira Aguiar} & Universidade Federal de Minas Gerais \\
\lattes{http://lattes.cnpq.br/0732210491984563}{Maria Emília Xavier Guimarães} & Universidade Federal Fluminense \\
\lattes{http://lattes.cnpq.br/2938214220427914}{Maria Eugênia Silva Nunes} & Universidade Federal de Ouro Preto \\
\lattes{ http://lattes.cnpq.br/6257416510668971}{Maria Fernanda Araujo de Resende} & Universidade Federal do ABC \\
\lattes{ http://lattes.cnpq.br/8777492109806507}{Maria Isabel Almeida de Oliveira} & Instituto Federal da Bahia \\
\lattes{http://lattes.cnpq.br/5146123472758388}{Maria Lewtchuk Espindola} & Universidade Federal da Paraíba \\
\lattes{http://lattes.cnpq.br/1901405515852408}{Maria Oswald Machado de Matos} & Pontifícia Universidade Católica do Rio de Janeiro \\
\lattes{ http://lattes.cnpq.br/8985268237852917}{Maria Simone Kugeratski Souza} & Universidade Federal de Santa Catarina \\
\lattes{http://lattes.cnpq.br/8627506263906532}{Maria Teresa Climaco Santos Thomaz} & Universidade Federal Fluminense \\
\lattes{http://lattes.cnpq.br/0925491332464897}{Mariana Malard Sales Andrade} & Universidade de Brasília \\
\lattes{http://lattes.cnpq.br/5641001357715339}{Mariana Rodrigues Barros} & Universidade Federal de Minas Gerais \\
\lattes{ http://lattes.cnpq.br/1990215293448663}{Mariana Zancan Tonel} & Universidade Franciscana \\
\lattes{http://lattes.cnpq.br/7364478567235458}{Marilia Junqueira Caldas} & Universidade de São Paulo \\
\lattes{ http://lattes.cnpq.br/2332971935511148}{Martine Chevrollier} & Universidade Federal Rural de Pernambuco \\
\lattes{ http://lattes.cnpq.br/6588189278676621}{Nadja Kolb Bernardes} & Universidade Federal de Pernambuco \\
\lattes{http://lattes.cnpq.br/6560563332731046}{Paula Homem-de-Mello} & Universidade Federal do ABC \\
\lattes{http://lattes.cnpq.br/2704717555047499}{Priscila Machado Vieira Lima} & Universidade Federal do Rio de Janeiro \\
\lattes{http://lattes.cnpq.br/3951800799770483}{Priscila Valdênia dos Santos} & Universidade Federal de Pernambuco \\
\lattes{http://lattes.cnpq.br/9476522311181916}{Raissa Fernandes Pessoa Mendes} & Universidade Federal Fluminense \\
\lattes{ http://lattes.cnpq.br/4474695838825056}{Regina Lélis de Sousa} & Universidade Federal do Norte do Tocantins \\
\lattes{http://lattes.cnpq.br/4788702298711499}{Regina Melo Silveira} & Universidade de São Paulo \\
\lattes{http://lattes.cnpq.br/3193435909206442}{Renata Zukanovich Funchal} & Universidade de São Paulo \\
\lattes{http://lattes.cnpq.br/5358286408119219}{Romarly Fernandes da Costa} & Universidade Federal do ABC \\
\lattes{http://lattes.cnpq.br/1439240219060111}{Sandra Denise Prado} & Universidade Federal do Rio Grande do Sul \\
\lattes{http://lattes.cnpq.br/0732996686411020}{Santosh Shelly Sharma} & Universidade Estadual de Londrina \\
\lattes{http://lattes.cnpq.br/3585326481456738}{Silvana Perez} & Universidade Federal do Pará \\
\lattes{http://lattes.cnpq.br/8148904322500734}{Silvania Alves de Carvalho} & Universidade Federal Fluminense \\
\lattes{http://lattes.cnpq.br/0527275499843168}{Silvete Coradi Guerini} & Universidade Federal do Maranhão \\
\lattes{http://lattes.cnpq.br/9883014782075093}{Simone Silva Alexandre} & Universidade Federal de Minas Gerais \\
\lattes{http://lattes.cnpq.br/5434389218137493}{Sonia Geraij Mokarzel} & Pontifícia Universidade Católica de São Paulo \\
\lattes{http://lattes.cnpq.br/8537174715205717}{Solange Binotto Fagan} & Universidade Franciscana \\
\lattes{http://lattes.cnpq.br/3532696501366872}{Tatiana Cardoso e Bufalo} & Universidade Federal de Lavras \\
\lattes{http://lattes.cnpq.br/7843291258244995}{Tatiana Gabriela Rappoport} & Universidade Federal do Rio de Janeiro\\
\lattes{ http://lattes.cnpq.br/8952649290540577}{Tereza Cristina da Rocha Mendes} & Universidade de São Paulo \\
\lattes{http://lattes.cnpq.br/8526560666304076}{Thaís Victa Trevisan} & Universidade de São Paulo \\
\lattes{ http://lattes.cnpq.br/2089952733406400}{Thereza Cristina de Lacerda Paiva} & Universidade Federal do Rio de Janeiro \\
\lattes{ http://lattes.cnpq.br/5008222023801977}{Vivian Vanessa França Henn} & Universidade Estadual Paulista \\
\lattes{http://lattes.cnpq.br/8613500501440021}{Zhanna Gennadyevna Kuznetsova} & Universidade Federal do ABC \\

\end{longtable}
\endgroup

\end{appendices}





\newpage
\bibliography{sn-bibliography}

\end{document}